\documentclass[aps,amsmath]{revtex4}
\usepackage{graphicx}  % needed for figures
\usepackage{dcolumn}   % needed for some tables
\usepackage{amssymb}   % for math
\usepackage{rotating}  % sideways tables
\def\MET{{\mbox{$E\kern-0.57em\raise0.19ex\hbox{/}_{T}$}}}
\def\met{{\mbox{$E\kern-0.57em\raise0.19ex\hbox{/}_{T}$}}}
\def\DZ{D0 }

\begin{document}
%\linenumbers

% remove the following for publication
%\rightline{Send comments to d0-run2eb-004@fnal.gov}
%\rightline{\& d0-run2eb-005@fnal.gov}
%\rightline{\& d0-run2eb-013@fnal.gov}
%\rightline{\& d0-run2eb-029@fnal.gov}
%\rightline{by March 9th, noon (FNAL time)}
%% remove the space for publication
\rightline{FERMILAB-CONF-11-413-E}
\rightline{CDF Note 10626}
\rightline{\DZ Note 6243}
\vskip0.5in

\title{Combined CDF and \DZ upper limits on Fermiophobic Higgs Boson Production with up to
8.2 fb{\boldmath $^{-1}$} of {\boldmath $p{\bar{p}}$} data\\[2.5cm]}

\author{%{\bf DRAFT--INTERNAL WORKING DOCUMENT for CDF and D0 Collaborations ONLY\\}
The TEVNPH Working Group\footnote{The Tevatron
New-Phenomena and Higgs Working Group can be contacted at
TEVNPHWG@fnal.gov. More information can be found at http://tevnphwg.fnal.gov/.}
 }
%\vskip*2cm
\affiliation{\vskip0.3cm for the CDF and \DZ Collaborations\\ \vskip0.2cm
\today}
%\vskip*2cm
\begin{abstract}
\vskip0.3in
We combine the results of searches by the CDF and D0 collaborations for a fermiophobic Higgs
boson ($H_f$) produced via the processes $WH_f$, $ZH_f$, and vector-boson fusion
in $p{\bar{p}}$ collisions at $\sqrt{s}=1.96$~TeV at the Fermilab Tevatron collider.
The analyses seek Higgs boson decays to $W^+W^-$ and $\gamma\gamma$.
With up to 8.2~fb$^{-1}$ of integrated luminosity analyzed at CDF and up to
8.2~fb$^{-1}$ at D0, we obtain a 95\% CL lower bound on the mass of the Higgs boson in the
fermiophobic Higgs model of 119~GeV/$c^2$.
\\[2cm]
{\hspace*{5.5cm}\em Preliminary Results}
\end{abstract}

\maketitle

\newpage
%\vspace*{0.1cm}
%%%%%%%%%%%%%%%%%%%%%%%%%%%%%%%%%%%%%%%%%%%%%%%%%%%%%%%%%%%%%%%%%%%%%%%
%%%%%%%%%%%%%%%%%%%%%%%%%%%%%%%%%%%%%%%%%%%%%%%%%%%%%%%%%%%%%%%%%%%%%%%
\section{Introduction} %%%%%%%%%%%%%%

The mechanism for the breaking of the $SU(2)\times U(1)$ electroweak gauge
symmetry~\cite{higgs} of the standard model (SM)~\cite{gws} 
has thus far evaded experimental verification
even though the SM accommodates current collider data and it has been tested with
remarkable precision.  Searches for the SM Higgs boson at LEP~\cite{lephiggs}, the Tevatron~\cite{tevsm},
and the LHC~\cite{atlassmhiggs,cmssmhiggs} have yet to confirm its existence.

The mechanism of electroweak symmetry breaking 
may offer a richer phenomenology than expected in the minimal version
of the SM.  Several Higgs bosons may exist, or the Higgs boson(s) may have couplings that are different
from those expected in the minimal model.  Here we explore the possibility that the lightest
Higgs scalar does not couple to fermions at tree level, but is in all other ways SM-like.
In order to provide masses to the fermions,
additional degrees of freedom must exist in the Higgs sector, and models with Higgs doublets
and triplets are possible~\cite{dtrip,Akeroyd:1995hg,Akeroyd:2010eg}.   We consider here
a model in which all particles and interactions beyond those of the SM have no phenomenological 
impact for observables at the Tevatron aside
from suppressing the fermion couplings to the Higgs boson.  We call this model the 
fermiophobic Higgs model (FHM).  In the following, we denote the fermiophobic Higgs boson $H_f$, the SM Higgs boson
$H_{\rm{SM}}$, and use the symbol $H$ in a context which is valid for both.

In the FHM, the production of Higgs bosons
at hadron colliders via the process $gg\rightarrow H_f$ is suppressed to a negligible rate.
In the SM, this process proceeds through one-loop intermediaries via fermion 
couplings to the Higgs boson, and at two or more loops 
via fermion and boson couplings to the Higgs boson.  Two-loop electroweak processes contribute
to $gg\rightarrow H_f$ production
in the fermiophobic Higgs model (FHM), but at a vanishingly small rate, as their main contribution
in the SM are through interference with one-loop processes~\cite{anastasiou}.  The associated production
mechanisms $p{\bar{p}}\rightarrow WH_f+X$ and $p{\bar{p}}\rightarrow ZH_f+X$, 
as well as the vector-boson-fusion (VBF)
processes $q^\prime{\bar{q}}\rightarrow H_fq^{\prime\prime}{\bar{q}}^{\prime\prime\prime}$ remain
nearly unchanged relative to the corresponding processes in the SM.
Table~\ref{tab:fpxsbr} lists the production rates
we use for the $WH_f$, $ZH_f$, and VBF modes.  We neglect the cross section for $gg\rightarrow H_f$ in this
model.   We use the $WH_{\rm{SM}}$ and $ZH_{\rm{SM}}$ cross sections from Ref.~\cite{djouadibaglio}, which are based on
the next-to-leading-order (NLO) calculation of
{\sc v2hv}~\cite{v2hv} and include next-to-next-to-leading-order (NNLO) 
QCD contributions~\cite{vhnnloqcd}, as well
as one-loop electroweak corrections~\cite{vhewcorr}.
We use the VBF cross section computed in the SM at NNLO in QCD in Ref.~\cite{vbfnnlo}.
Electroweak corrections to the VBF production cross section are computed
with the {\sc hawk} program~\cite{hawk}, and are small and negative (2-3\%)
in the Higgs boson mass range considered here.  We include these corrections in the VBF
cross sections used in this combination of results from the Tevatron.  We use the systematic uncertainties
on the production cross section predictions from their respective publications, treating the $WH_f$ and
$ZH_f$ production cross section uncertainties as fully correlated with each other, and uncorrelated with the
VBF production cross section uncertainty.

Another consequence of the FHM is to modify the decay branching ratios of the Higgs boson
with respect to the SM predictions.  Direct decays to fermions are forbidden; decays to fermions can
proceed only via loops involving weak gauge bosons, and these are highly suppressed due to a spin flip
required in the loop.  The partial width of the Higgs boson decay to a pair of gluons, which in the SM
is comparable to that of the decay to a pair of charm quarks, is negligibly small in the FHM.  The remaining
decays, to $\gamma\gamma$, $W^+W^-$, $Z\gamma$, and $ZZ$, account for nearly the entire decay width.  For
Higgs boson masses below twice the $W$ boson mass, $m_{H_f}\ll 2M_W$, the favored direct 
decay to $W^+W^-$ is reduced because one or both of the $W$ bosons is far from resonance.  Nonetheless, because
the competing decay modes have small partial widths, the branching to $W^+W^-$ is still
dominant, particularly for a heavier Higgs boson with $m_{H_f}>120$~GeV/$c^2$.  The branching fraction
${\cal{B}}(H_f\rightarrow\gamma\gamma$) is greatly enhanced over ${\cal{B}}(H_{\rm{SM}}\rightarrow\gamma\gamma$) for all $m_{H}$,
and its clean signature and excellent mass resolution
provide most of the search sensitivity for $m_{H_f}<120$~GeV/$c^2$.

Table~\ref{tab:fpxsbr}
lists the decay branching fractions we assume in the searches described below, all of which were computed
with {\sc hdecay}~\cite{hdecay}.  The dominant uncertainties
on the SM Higgs branching fractions in the range 100~GeV$/c^2<m_{H_{\rm{SM}}}<$~200~GeV$/c^2$
arise from uncertainties on the $b$ quark mass $m_b$, the charm quark mass $m_c$, and 
the strong coupling constant $\alpha_s$~\cite{dblittlelhc,lhcbrunc}.  
Because these uncertainties
affect only the decay modes in which a fermion (specifically, a quark) couples to the Higgs boson, they
do not affect the Higgs boson's branching ratios in the FHM, and therefore they do not contribute to this
analysis.

The CDF and D0 collaborations have searched for the SM Higgs boson in  
$H_{\rm{SM}}\rightarrow\gamma\gamma$ decays~\cite{cdfhggsm,d0hggsmfp}.   These searches cannot be directly reinterpreted
as constraints on the FHM by simply scaling the cross sections and branching ratios,
because the kinematic distributions of the Higgs bosons, their decay products, and the particles produced
in association with the Higgs boson differ between the FHM and the SM.
These differences arise from the absence of $gg\rightarrow H_f$ production in the FHM,
and the increased fraction of the presence of an associated vector 
boson $W$ or $Z$, or recoiling quark jets in the case of VBF.
The transverse momentum ($p_T$) spectrum of the Higgs boson in the FHM is thus much harder than it is in the SM, affecting the
signal acceptance and the ability to separate the signal from the backgrounds.
The $H_f\rightarrow\gamma\gamma$ analyses have therefore been reoptimized for the FHM, taking advantage of the
higher $p_T$ of the Higgs boson~\cite{cdfhggfp,d0hggsmfp}.  These analyses are updates of previous searches
for the Higgs boson in the FHM using 3.0~fb$^{-1}$ of data at CDF~\cite{prevcdffhm} and 2.7~fb$^{-1}$ 
of data at D0~\cite{prevd0fhm}.
The CMS collaboration has also sought Higgs bosons decaying to photon pairs, and sets a lower limit 
on $m_{H_f}$ in the FHM of 112~GeV/$c^2$ at the 95\% CL~\cite{cmsdipho}.  ATLAS has sought the SM Higgs boson in the
decay $H_{\rm{SM}}\rightarrow\gamma\gamma$ in 1.1 fb$^{-1}$ of data~\cite{atlassmhiggs}.  The four LEP collaborations,
ALEPH, DELPHI, L3, and OPAL, have searched for $e^+e^-\rightarrow ZH_f\rightarrow Z\gamma\gamma$, and from their
combined results report a lower limit on $m_{H_f}$ of 108.2~GeV/$c^2$ in the FHM at the 95\% CL~\cite{lepdipho}.

\begin{table}
\begin{center}
\caption{
The production cross sections for $WH_f$, $ZH_f$, and VBF at NNLO  Also listed are
decay branching fractions $({\cal{B}})$ 
of the Higgs boson in the Fermiophobic Higgs model computed with {\sc hdecay}~\protect{\cite{hdecay}}.
}
\label{tab:fpxsbr}
\vspace{0.3cm}
\begin{tabular}{|c|c|c|c|c|c|c|}\hline
$m_{H_f}$ (GeV/$c^2$) & $\sigma_{WH}$ (fb) & $\sigma_{ZH}$ (fb) & $\sigma_{\rm{VBF}}$ (fb) & 
   ${\cal{B}}(H_f\rightarrow\gamma\gamma)$ &
   ${\cal{B}}(H_f\rightarrow W^+W^-)$ & 
   ${\cal{B}}(H_f\rightarrow ZZ)$ \\ \hline
100 & 291.9 & 169.8 & 97.2 &   0.185      & 0.735  &     0.0762     \\ 
105 & 248.4 & 145.9 & 89.7 &   0.104      & 0.816  &     0.0733     \\ 
110 & 212.0 & 125.7 & 82.7 &   0.0603     & 0.853  &     0.0788     \\ 
115 & 174.5 & 103.9 & 76.4 &   0.0366     & 0.866  &     0.0887     \\ 
120 & 150.1 &  90.2 & 70.7 &   0.0233     & 0.869  &     0.0993     \\ 
125 & 129.5 &  78.5 & 65.3 &   0.0156     & 0.868  &     0.109      \\ 
130 & 112.0 &  68.5 & 60.4 &   0.0107     & 0.867  &     0.116      \\ 
135 &  97.2 &  60.0 & 55.9 &   0.759$\times 10^{-2}$  & 0.866  &     0.120  \\ 
140 &  84.6 &  52.7 & 51.8 &   0.544$\times 10^{-2}$  & 0.868  &     0.121  \\ 
145 &  73.7 &  46.3 & 48.1 &   0.390$\times 10^{-2}$  & 0.874  &     0.118  \\ 
150 &  64.4 &  40.8 & 44.6 &   0.273$\times 10^{-2}$  & 0.886  &     0.108  \\ 
155 &  56.2 &  35.9 & 41.2 &   0.176$\times 10^{-2}$  & 0.909  &     0.0871  \\ 
160 &  48.5 &  31.4 & 38.2 &   0.835$\times 10^{-3}$  & 0.951  &     0.0466  \\ 
165 &  43.6 &  28.4 & 36.0 &   0.334$\times 10^{-3}$  & 0.975  &     0.0236  \\ 
170 &  38.5 &  25.3 & 33.4 &   0.226$\times 10^{-3}$  & 0.975  &     0.0246  \\ 
175 &  34.0 &  22.5 & 31.0 &   0.179$\times 10^{-3}$  & 0.966  &     0.0332  \\ 
180 &  30.1 &  20.0 & 28.8 &   0.148$\times 10^{-3}$  & 0.939  &     0.0609  \\ 
185 &  26.9 &  17.9 & 26.9 &   0.118$\times 10^{-3}$  & 0.848  &     0.152   \\ 
190 &  24.0 &  16.1 & 25.0 &   0.979$\times 10^{-4}$  & 0.788  &     0.212   \\ 
195 &  21.4 &  14.4 & 23.3 &   0.852$\times 10^{-4}$  & 0.759  &     0.241   \\ 
200 &  19.1 &  13.0 & 21.6 &   0.759$\times 10^{-4}$  & 0.742  &     0.258   \\ \hline
\end{tabular}		    	   
\end{center}		    
\end{table}

In addition to the searches for $H_f\rightarrow\gamma\gamma$, we also combine CDF's
and D0's searches for $H\rightarrow W^+W^-$, taking advantage of its enhanced 
branching fraction in the FHM.  
As for $H\rightarrow\gamma\gamma$, the searches for $H_{\rm{SM}}\rightarrow W^+W^-$ 
cannot be interpted directly in the FHM due to the different mixture of production modes.

The CDF collaboration's $H_{\rm{SM}}\rightarrow W^+W^-$ analyses~\cite{cdfwwsum11} keep separate account of the
predictions from $gg\rightarrow H_{\rm{SM}}$, $WH_{\rm{SM}}$, $ZH_{\rm{SM}}$, and VBF in each of the contributing channels.
The distributions of the neural-network discriminants
of these analyses are reinterpreted in the FHM by setting the $gg\rightarrow H_f$ component to zero
and by scaling the remaining signal components by the ratio of branching ratio predictions 
${\cal{B}}(H_f\rightarrow W^+W^-)/{\cal{B}}(H_{\rm{SM}}\rightarrow W^+W^-)$.
CDF's searches for opposite-charge dilepton events in the final state
$H_{\rm{SM}}\rightarrow W^+W^-\rightarrow \ell^+\nu_{\ell}\ell^{\prime-}{\bar{\nu}}_{\ell^\prime}$ (where $\ell,\ell^\prime=e,\mu$),
are separated into categories based on the number of reconstructed jets accompanying the leptons and missing
transverse energy in the event.  The $gg\rightarrow H$ process, which contributes negligibly in the FHM,
most frequently produces a Higgs boson without additional jets, while the $WH$, $ZH$, and $VBF$ production modes more commonly
produce Higgs bosons with additional jets.  This division of events into jet categories optimizes the search in the FHM
and does not require developing a separate set of analysis channels, unlike the case of $H_f\rightarrow\gamma\gamma$.
The opposite-charge low-$m_{\ell\ell}$ channel and the channels seeking $H\rightarrow W^+W^-$, in which one
$W$ decays into a $\tau$ lepton which then decays to hadrons$+\nu_\tau$, while the other $W$ decays to $e\nu$ or $\mu\nu$, are
included in the combination.  The trilepton and like-charge dilepton analyses,
also included in CDF's SM $H\rightarrow W^+W^-$ search~\cite{cdfwwsum11}, are included here as well, as they
are targeted at the $WH$ and $ZH$ signal contributions.

The D0 SM searches for $H_{\rm{SM}}\rightarrow W^+W^-$ included here are the like-charge dilepton searches~\cite{dzWWW}, targeting the
$WH$ and $ZH$ production modes, in which the $W$ or $Z$ produced in association with the Higgs boson
decays leptonically and forms a like-charge pair with a lepton from one of the $W$ decays.  They are reinterpteted here
as searches in the FHM by scaling by the ratio of branching ratios in the SM and FHM.  CDF uses neural-network discriminants
and D0 uses boosted decision trees as the final step in separating a possible signal from the backgrounds.
  The same multivariate discriminant functions used in the SM searches are used here
without reoptimization for the FHM.  A summary of the included channels and their luminosities is provided in 
Table~\ref{tab:channels}.

\begin{table}
\caption{\label{tab:channels}Luminosity, explored mass range and references
for the different processes and final states ($\ell$ = $e$ or $\mu$) for
the CDF analyses.  The label ``$2\times$'' refers to a separation based on lepton categories.}
\begin{tabular}{|lccc|} \hline
Channel & Luminosity  & $m_{H_f}$ range & Reference \\
        & (fb$^{-1}$) & (GeV/$c^2$) &           \\ \hline
CDF $H_f \rightarrow \gamma \gamma$ \ \ \  % (CC,CP,CC-Conv,PC-Conv)  
                                                                     & 7.0  & 100-150 & \cite{cdfhggfp} \\
CDF $H\rightarrow W^+ W^-$ \ \ \ 2$\times$(0 jets,1 jet)+(2 or more jets)+(low-$m_{\ell\ell}$)+($e$-$\tau_{\rm{had}}$)+($\mu$-$\tau_{\rm{had}}$) & 8.2  & 110-200 & \cite{cdfwwsum11} \\
CDF $WH \rightarrow WW^+ W^-$ \ \ \ (same-sign leptons)+(tri-leptons)                                                                  & 8.2  & 110-200 & \cite{cdfwwsum11} \\
CDF $ZH \rightarrow ZW^+ W^-$ \ \ \ (tri-leptons with 1 jet)+(tri-leptons with 2 or more jets)                                         & 8.2  & 110-200 & \cite{cdfwwsum11} \\
D0 $H_f \rightarrow \gamma \gamma$                                 & 8.2  & 100-150 & \cite{d0hggsmfp} \\
D0 $VH \rightarrow \ell^\pm \ell^\pm\ + X $ \ \ \  & 5.3  & 115-200 & \cite{dzWWW} \\ \hline
\end{tabular}
\end{table}

 We use the same statistical methods employed in
Ref.~\cite{tevwwprl}, namely the modified frequentist (CL$_{\rm s}$)
and Bayesian techniques, in order to evaluate the
results.  Pseudo-experiments drawn from the background predictions varied according to their
systematic uncertainties are used to compute
the limits we expect to obtain in the absence of signal.  Correlated systematic
uncertainties are treated in the same way as they are in
Ref.~\cite{tevwwprl}.  The sources of correlated uncertainty between
CDF and D0 are the total inelastic $p\bar{p}$ cross section used in
the luminosity measurement, the SM diboson background production cross
sections ($WW$, $WZ$, and $ZZ$), and the $t{\bar{t}}$ and single top
quark production cross sections.
Instrumental effects such as trigger efficiencies, photon and lepton
identification efficiencies and misidentification rates, and the jet
energy scales used by CDF and D0 remain uncorrelated.  To minimize the
degrading effects of systematics on the sensitivity of the search, the signal
and background contributions are fit to data by
maximizing a likelihood function over the systematic uncertainties for
both the background-only and signal+background
hypotheses~\cite{fitting}.  We include the theoretical uncertainties on the predictions of the
production cross sections when quoting limits normalized to them and
when quoting our limit on $m_{H_f}$ in the FHM.

\begin{figure}
 \begin{center}
\includegraphics[width=0.8\textwidth]{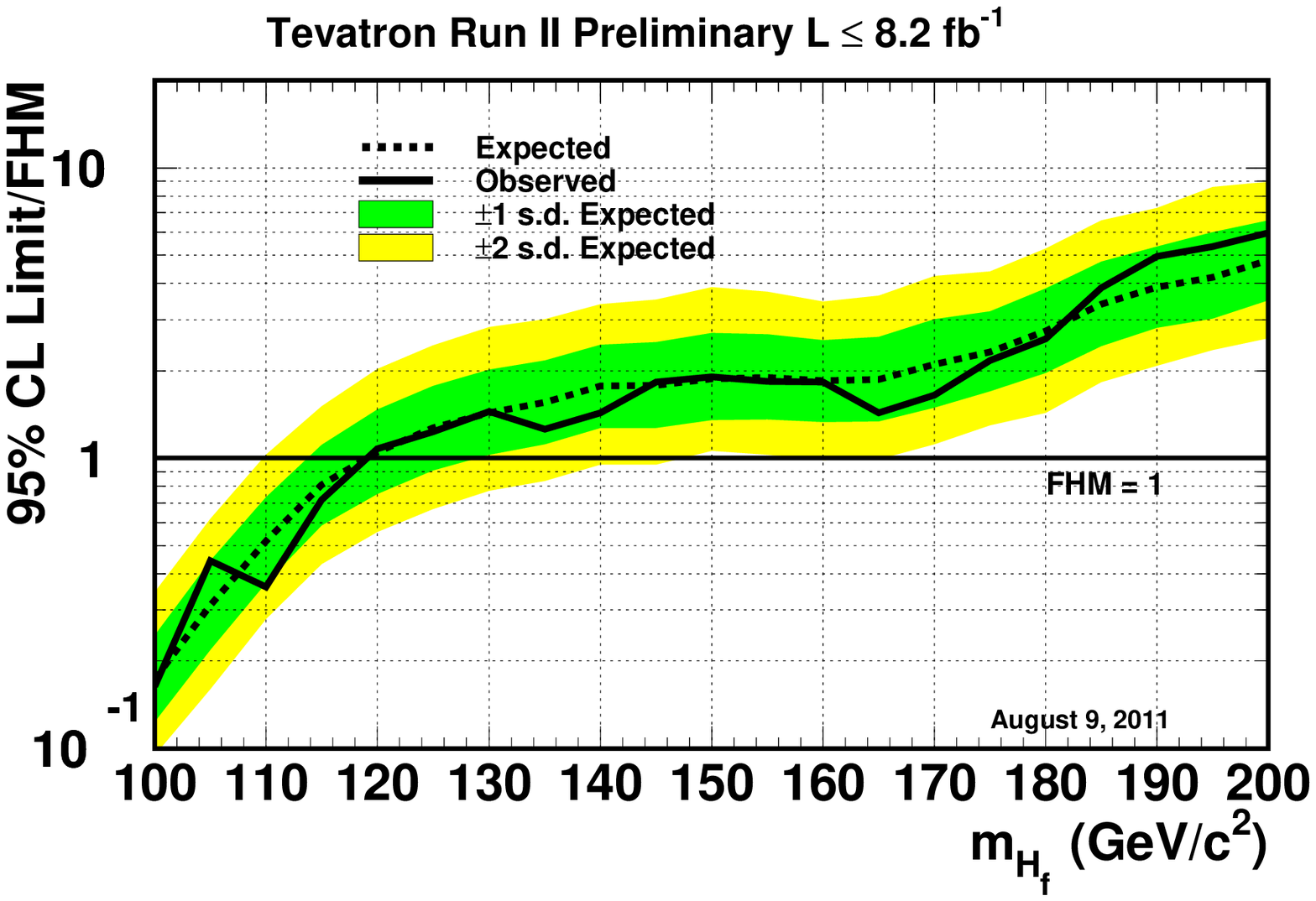}
 \end{center}
 \caption{
 \label{fig:limits}  Observed (solid black line) and median expected (dashed
black line) 95\% CL upper limits from the Tevatron on Higgs boson production in the fermiophobic Higgs model (FHM).
 The shaded bands indicate the $\pm 1$~standard deviation (s.d.) and $\pm 2$~s.d. intervals
on the distribution of the limits that are expected in the absence of a contribution from the Higgs boson.    
}
 \end{figure}

 The combined limits on Higgs boson production normalized to predictions of the FHM 
are listed in Table~\ref{tab:limits}
for both the CL$_{\rm s}$ and the Bayesian methods, for 
100~GeV/$c^2<m_{H_f}<200$~GeV/$c^2$, in 5~GeV/$c^2$ steps.  
The expected and observed limits for both methods agree
within 5\% for nearly all $m_{H_f}$ values except for $m_{H_f}>185$~GeV/$c^2$, where
the agreement still remains close for the expected limit and is within 10\% for the observed limits.
We choose to present the limits calculated with the Bayesian approach which was selected
{\it a priori}.  The final limits are shown in Fig.~\ref{fig:limits}.  To quote a limit on $m_{H_f}$ in the FHM,
we interpolate the limits linearly between the sampled values of $m_{H_f}$ and report the locations
at which the observed and expected limit functions cross unity.  We exclude $m_{H_f}<119$~GeV/$c^2$ at
the 95\% CL, and the expected exclusion range is also $m_{H_f}<119$~GeV/$c^2$.

 \begin{table}
 \begin{center}
 \caption{\label{tab:limits}  Observed and expected limits from the combined Tevatron search for Higgs boson
production in the fermiophobic Higgs model (FHM)}
 \begin{tabular}{|c|cc|cc|}\cline{2-5}
\multicolumn{1}{c|}{} & \multicolumn{2}{c|}{Bayesian} & \multicolumn{2}{c|}{CL$_{\rm{s}}$} \\ \hline
 $m_{H_f}$ & obs & Median exp & obs & Median exp \\
 (GeV/$c^2$) & (Limit/FHM) & (Limit/FHM) & (Limit/FHM) & (Limit/FHM) \\
 \hline
100 &     0.16 &         0.17 & 0.17	&	0.17    \\
105 &     0.44 &         0.31 & 0.44	&	0.31    \\
110 &     0.36 &         0.52 & 0.38	&	0.52    \\
115 &     0.72 &         0.81 & 0.75	&	0.81    \\
120 &     1.07 &         1.05 & 1.06	&	1.03    \\
125 &     1.23 &         1.27 & 1.20	&	1.22    \\
130 &     1.45 &         1.44 & 1.47	&	1.41    \\
135 &     1.26 &         1.55 & 1.29	&	1.51    \\
140 &     1.44 &         1.77 & 1.47	&	1.69    \\
145 &     1.83 &         1.78 & 1.80	&	1.75    \\
150 &     1.90 &         1.88 & 1.87	&	1.83    \\
155 &     1.83 &         1.90 & 1.80    &       1.95    \\
160 &     1.83 &         1.84 & 1.78    &       1.86    \\
165 &     1.43 &         1.87 & 1.46    &       1.85    \\
170 &     1.64 &         2.11 & 1.66    &       2.09    \\
175 &     2.17 &         2.32 & 2.11    &       2.36    \\
180 &     2.57 &         2.75 & 2.51    &       2.73    \\
185 &     3.85 &         3.39 & 3.66    &       3.34    \\
190 &     4.97 &         3.89 & 4.48    &       3.90    \\
195 &     5.36 &         4.20 & 4.90    &       4.36    \\
200 &     5.96 &         4.80 & 5.34    &       4.85    \\
 \hline				 
 \end{tabular}
 \end{center}
 \end{table}

In summary, we present a combination of CDF and D0 searches for the Higgs boson in a model in which
the tree-level couplings of the Higgs boson to fermions vanish, while retaining standard model couplings to bosons.
We exclude, at the 95\% CL, a Higgs boson of mass $m_{H_f}<119$~GeV/$c^2$ in this model.  This is the most restrictive
limit to date on the fermiophobic Higgs model.

\begin{center}
{\bf Acknowledgements}
\end{center}

We thank the Fermilab staff and the technical staffs of the
participating institutions for their vital contributions. 
This work was supported by  
DOE and NSF (USA),
CONICET and UBACyT (Argentina), 
CNPq, FAPERJ, FAPESP and FUNDUNESP (Brazil),
CRC Program, CFI, NSERC and WestGrid Project (Canada),
CAS and CNSF (China),
Colciencias (Colombia),
MSMT and GACR (Czech Republic),
Academy of Finland (Finland),
CEA and CNRS/IN2P3 (France),
BMBF and DFG (Germany),
Ministry of Education, Culture, Sports, Science and Technology (Japan), 
World Class University Program, National Research Foundation (Korea),
KRF and KOSEF (Korea),
DAE and DST (India),
SFI (Ireland),
INFN (Italy),
CONACyT (Mexico),
NSC(Republic of China),
FASI, Rosatom and RFBR (Russia),
Slovak R\&D Agency (Slovakia), 
Ministerio de Ciencia e Innovaci\'{o}n, and Programa Consolider-Ingenio 2010 (Spain),
The Swedish Research Council (Sweden),
Swiss National Science Foundation (Switzerland), 
FOM (The Netherlands),
STFC and the Royal Society (UK),
and the A.P. Sloan Foundation (USA).

\end{document}